\documentclass[11pt, a4paper]{article}
\pdfoutput=1 

\usepackage[height=8.85in,width=6.75in]{geometry}

\usepackage{graphicx,rotating}     
\usepackage[bookmarksopen,colorlinks=true,linkcolor=steelblue,
citecolor=orangered,urlcolor=darkred,linktoc=all]{hyperref}

\usepackage{amsmath,amssymb}
\usepackage{slashed}
\usepackage{bm}
\usepackage{bbm}
\usepackage{cite}
\usepackage{comment}
\usepackage[utf8]{inputenc}
\usepackage{accents}
\usepackage[footnotesize]{caption}
\usepackage{subcaption}
\usepackage{cancel}

\newcommand{\abs}[1]{\left\vert {#1} \right\vert}
\newcommand{\hyphen}{\mathchar`-\mathchar`-}

\usepackage{xcolor}
\definecolor{darkred}{rgb}{0.7, 0., 0.}
\definecolor{orangered}{rgb}{1,0.27,0.}
\definecolor{steelblue}{rgb}{0.275,0.51, 0.706}
\definecolor{forestgreen}{rgb}{0.13,0.55,0.13}

\usepackage{tikz}
\usepackage{tikz-feynman}
\tikzfeynmanset{compat=1.0.0}
\pgfdeclarelayer{bg}    
\pgfsetlayers{bg,main}  

\begin{document}


\begin{center}

\hfill UMN-TH-4127/22 \\
\hfill FTPI-MINN-22-18 \\

\vskip 0.5in

{\Huge \bfseries Reevaluation of heavy-fermion-induced\\ \vspace{0.3cm}
electron EDM at three loops
} \\
\vskip .8in

{\Large Yohei Ema$^{a}$, Ting Gao$^{b}$, Maxim Pospelov$^{a,b}$}

\vskip .3in
\begin{tabular}{ll}
$^a$& \!\!\!\!\!\emph{William I. Fine Theoretical Physics Institute, School of Physics and Astronomy,}\\[-.3em]
& \!\!\!\!\!\emph{University of Minnesota, Minneapolis, MN 55455, USA} \\
$^b$& \!\!\!\!\!\emph{School of Physics and Astronomy, University of Minnesota, Minneapolis, MN 55455, USA}
\end{tabular}

\end{center}
\vskip .6in

\begin{abstract}
\noindent
Motivated by improved limits on electric dipole moments (EDMs), we revisit the three-loop light-by-light mechanism that transmits $CP$ violation from a heavy fermion to a light one. 
Applying it to the electron EDM induced by EDMs of heavier flavors, we find additional contributions at the order $ m_e/m_\mu $ or $m_e/m_\tau$ that were missing before and that make the result 1.4 times larger. 
Consequently we improve bounds on the tau EDM, and update indirect limits on the muon EDM as well as charm and bottom quark color EDMs.
\end{abstract}

\section{Introduction}
Searches for electric dipole moments (EDMs) have been a powerful probe of fundamental symmetry violation since the first proposal by Purcell and Ramsey in the 1950s\cite{Purcell:1950zz}. While a nonzero EDM of elementary particles has not yet been detected, one can expect new EDM experiments to continue improving sensitivity, eventually reaching benchmark values suggested by the Standard Model (SM) (see {\em e.g.} recent reevaluation in Ref. \cite{Ema:2022yra}). This situation makes EDM experiments sensitive to new physics, and specifically to new sources of CP violation beyond the SM.

In the lepton sector, the electron EDM has been in the focus of experimental and theoretical attention for several decades. The stable nature of electrons makes it possible to probe its EDM directly in paramagnetic atoms and molecules\cite{ACME:2018yjb}. There is also rising interest to muon EDM due to the muon $ g-2 $ anomaly~\cite{Crivellin:2018qmi,Aoyama:2020ynm,Muong-2:2021ojo}. Muon EDM can be directly measured in storage ring experiments\cite{Semertzidis:1999kv,Muong-2:2008ebm,Iinuma:2016zfu,Abe:2019thb,Adelmann:2021udj}, 
and indirectly constrained \cite{Ema:2021jds} using atomic and molecular EDM experiments~\cite{Graner:2016ses,ACME:2018yjb}. 

Direct measurement of tau-lepton EDM is difficult due to its short lifetime, so the knowledge of tau EDM mostly comes from measurements of tau pair-production.
At present the best constraint is inferred from the CP-violating effects in $ e^+e^-\to\tau^+\tau^- $ at the Belle Collaboration\cite{Belle:2002nla}
\begin{equation}
\begin{aligned}
    &\text{Re}(d_\tau)=\left( 1.15\pm 1.70\right)\times 10^{-17}e\,\text{cm},\\
    &\text{Im}(d_\tau)=\left( -0.83\pm 0.86\right)\times 10^{-17}e\,\text{cm}.
\end{aligned}
\end{equation}
More details on the current status of tau EDM search can be found in~\cite{Bernreuther:2021elu} 
and references therein.

A non-zero heavy fermion EDM, through loops, can induce EDMs of lighter fermions. 
It is first shown by Grozin, Khriplovich, and Rudenko in \cite{Grozin:2008nw} that heavy leptons (muon and tau lepton) can induce electron EDM at $ \alpha_{\text{EM}}^3 $ order. (Here the subscript EM stands for electromagnetic.) In this way \cite{Grozin:2008nw} puts indirect constraints on heavy lepton EDM based on electron EDM experiments. The purpose of this paper is to revisit this calculation and to correct an omission made in the previous work \cite{Grozin:2008nw}. Specifically, we will show that there is a contribution from the same set of diagrams in \cite{Grozin:2008nw} that was missing before, and this new term enhances the previous result by $  \sim 40\% $. Based on this updated calculation and the improved precision of paramagnetic EDM experiments, we provide a new indirect limit on tau EDM. As a by-product of this calculation, we also update our previous constraints on muon EDM~\cite{Ema:2021jds} and provide constraints on heavy quark color EDM (CEDM) 
based on light quark EDM.

\section{Evaluation of EDM light-by-light diagrams}
\begin{figure}[t]
    	\centering
    	\includegraphics[width=0.65\linewidth]{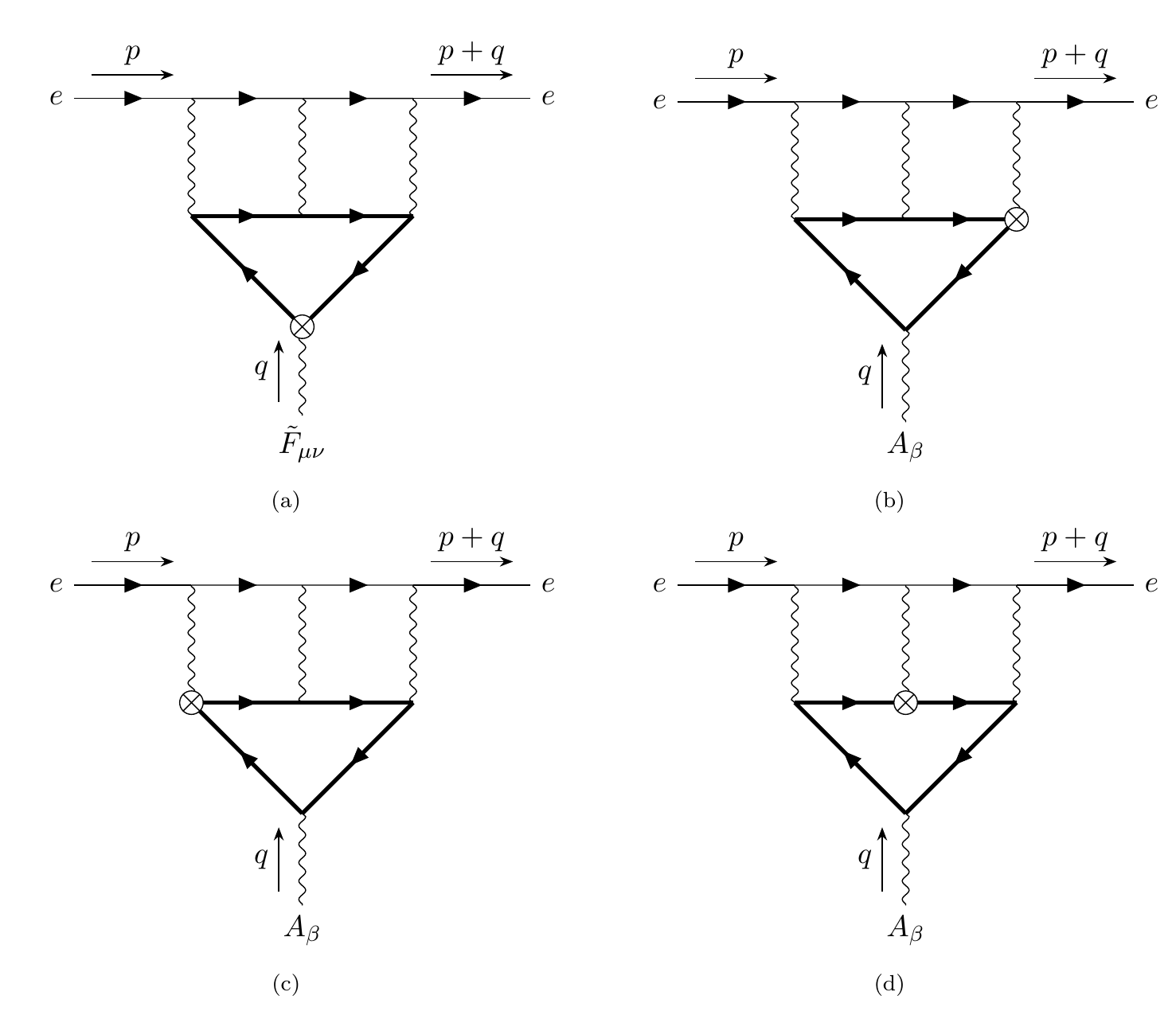}
		\caption{Examples of three-loop QED diagram for heavy lepton EDM contribution to electron EDM. The upper fermion line represents the electron and the lower fermion loop is formed by the heavy lepton. The crossed dot is the EDM vertex and replaces one of the 4 regular EM vertices on the heavy lepton loop.
		 The three photon lines connecting the heavy lepton loop and electron line have 6 possible permutations.}
		\label{3-loop QED diagram}
\end{figure}

In this section we reevaluate the heavy lepton induced electron EDM at three-loop. A total number of 24 diagrams contributes, including the diagrams in Fig.~\ref{3-loop QED diagram} and their permutations.

\begin{figure}[t]
    	\centering
    	\includegraphics[width=0.65\linewidth]{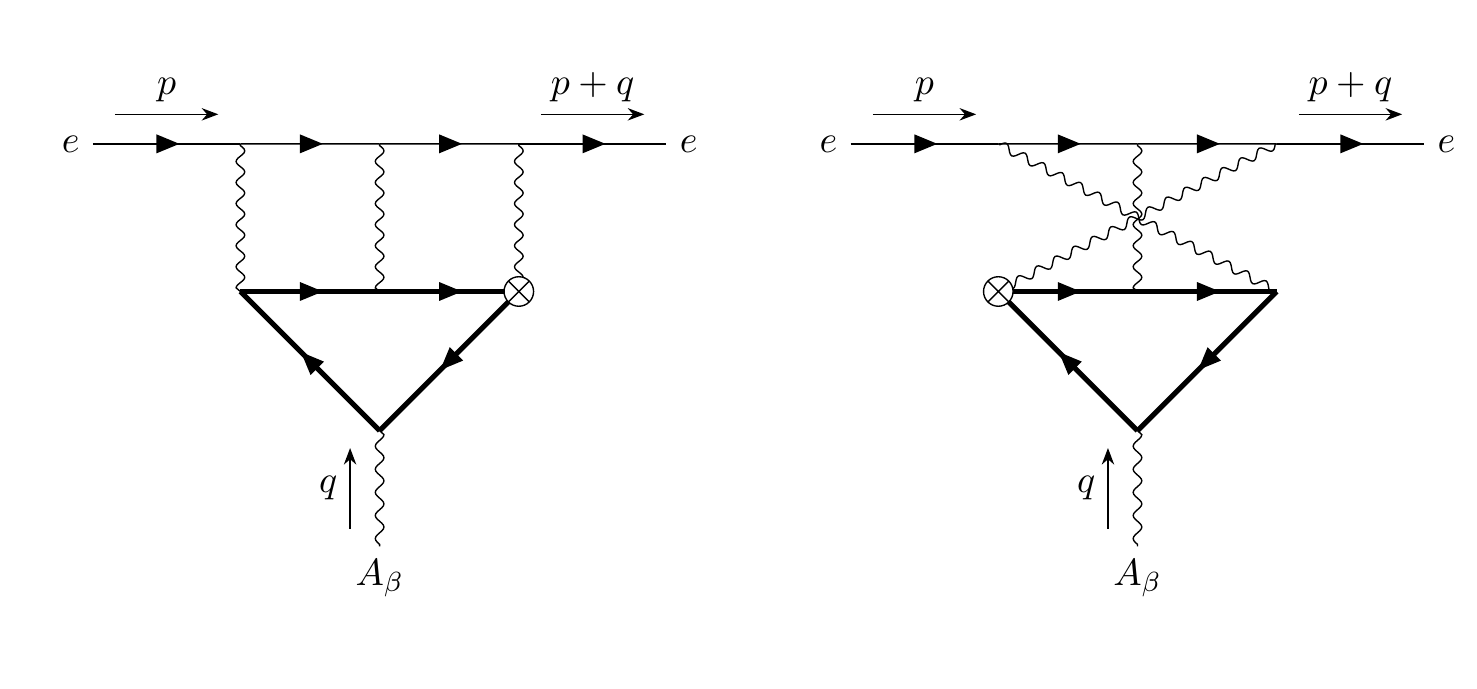}
		\caption{An example of related diagrams.}
		\label{equivalent diagram example}
\end{figure}

To get the EDM operator we expand the amplitude up to linear order in the external photon momentum $ q $. While full expression for arbitrary mass hierarchy can be found following \cite{Laporta:1992pa}, in practice we would like to explore the smallness of electron mass compared to a fermion mass inside the closed loop. Thus, observing that $ m_l/m_L\ll 1 $, where $ l=e $ refers to the electron, and $ L=\mu,\tau $ refers to heavy leptons, we evaluate the amplitude up to leading order in $ m_l/m_L $. Noticing that on account of Dirac equation $\left(\slashed{p}-m_l\right)l(p)=0 $, the expansion in $ m_l $ needs to be accompanied with an expansion in $ p $. So most generally, after the expansion one would get an expression of the following form:
\begin{equation}
    \label{EDM from Feynman diagram}
    \mathcal{M}=-i m_l q^\alpha A^\beta \bar{l}S_{\alpha\beta}^{(1)}l-i p^\rho q^\alpha A^\beta \bar{l}S_{\alpha\beta\rho}^{(2)}l.
\end{equation}
This expression contains two structures. While it is possible to convert one structure to the other with the help of the Dirac equation, we will keep both structures for later clarity, in order to compare our result
with the previous one.
The EDM operator can be written as
\begin{equation}
\label{EDM from Lagrangian}
    \mathcal{M}_{\text{EDM}}=-\frac{id_l}{2}\epsilon_{\mu\nu\alpha\beta} q^\alpha A^\beta \bar{l} \sigma^{\mu\nu}l= -\frac{id_l}{4m_l}p^\rho \epsilon_{\mu\nu\alpha\beta} q^\alpha A^\beta\bar{l} \{\sigma^{\mu\nu},\gamma_\rho\}l,
\end{equation}
where $q$ is the incoming photon momentum as in Fig.~\ref{3-loop QED diagram}.
Comparing Eqs.~(\ref{EDM from Feynman diagram}) and~(\ref{EDM from Lagrangian}), we see $ S_{\alpha\beta}^{(1)}\propto \epsilon_{\mu\nu\alpha\beta}\sigma^{\mu\nu}  $ and $ S_{\alpha\beta\rho}^{(2)}\propto \epsilon_{\mu\nu\alpha\beta}\{\sigma^{\mu\nu},\gamma_\rho\}  $. Using completeness of Dirac matrices and properties of the Levi-Civita tensor, we find
\begin{equation}
\label{projector}
    \begin{aligned}
        &S_{\alpha\beta}^{(1)}=-\frac{1}{8d(d-1)(d-2)(d-3)}\text{Tr}\left[S_{\gamma\delta}^{(1)}\epsilon^{\gamma\delta\kappa\lambda}\sigma_{\kappa\lambda}\right]\times \epsilon_{\mu\nu\alpha\beta}\sigma^{\mu\nu}, \\
        &S_{\alpha\beta\rho}^{(2)}=-\frac{1}{32d(d-1)(d-2)^2(d-3)}\text{Tr}\left[S_{\gamma\delta\eta}^{(2)}\epsilon^{\gamma\delta\kappa\lambda}\{\sigma_{\kappa\lambda},\gamma^{\eta}\}\right]\times \epsilon_{\mu\nu\alpha\beta}\{\sigma^{\mu\nu},\gamma_\rho\}.
    \end{aligned}
\end{equation}
This expression is generalized to arbitrary dimension $d$ for the later using inside the dimensionally-regularized loop expressions.
By writing $ S_{\alpha\beta}^{(1)} $ and $ S_{\alpha\beta\rho}^{(2)} $ in the form of Eq.~(\ref{projector}), we can focus on the scalar integral inside the trace, rather than a complicated integral with open tensor and Dirac indices. The scalar vacuum integral has a simple topology described by $ B_M $ in \cite{Broadhurst:1991fi} and can be reduced to two master integrals by repeated use of integration by parts \cite{Avdeev:1995eu,Steinhauser:2000ry,Chetyrkin:1981qh,Tkachov:1981wb,Martin:2016bgz}. One of the master integral is simply the product of three one-loop integrals, and the other is a three-loop integral corresponding to $ \textbf{E}(0,0,x,x) $ in \cite{Martin:2016bgz}. We use the \texttt{FIRE6} \cite{Smirnov:2019qkx} package to perform the integration-by-parts reduction and use the analytical expressions of master integrals in~\cite{Martin:2016bgz}. Divergences and gauge dependencies in the two structures cancel out separately and leave with us a finite result. Using the same notation as \cite{Grozin:2008nw}, we get
\begin{equation}
    \Delta d_l=a \frac{m_l}{m_L}\left(\frac{\alpha}{\pi}\right)^3 d_L,
\end{equation}
with
\begin{equation}
    \begin{aligned}
        & a^{(1)}_1=\frac{3}{2}\zeta(3)-\frac{19}{12} , \quad a^{(1)}_2=\frac{9}{4}\zeta(3)-1 , \quad a^{(1)}=a^{(1)}_1+a^{(1)}_2=\frac{15}{4}\zeta(3)-\frac{31}{12}\approx 1.924, \\
        & a^{(2)}_1=\frac{1}{2}\zeta(3)-\frac{1}{6} , \quad a^{(2)}_2=\frac{1}{2}\zeta(3)-\frac{5}{24} , \quad a^{(2)}=a^{(2)}_1+a^{(2)}_2=\zeta(3)-\frac{3}{8}\approx 0.827, \\
        & a=a^{(1)}+a^{(2)}=\frac{19}{4}\zeta(3)-\frac{71}{24}\approx 2.751,
    \end{aligned}
\end{equation}
where $ \zeta(s) $ is the Riemann zeta function with $ \zeta(3)\approx 1.202 $.
The contribution from expansion in $m_l/m_L$ is labeled by the upper index $``(1)"$, and the contribution from expansion in $p/m_L$ is labeled by the upper index $``(2)"$. The lower index indicates which set of diagram the result comes from, with $``1"$ representing the diagram (a) (and its permutations), and $``2"$ representing the sum of diagrams (b), (c), (d) (and their permutations). Note that the correct expression for the (a)-type of diagrams with a heavy quark EDM source and three intermediate gluons have been reported by us in our previous publication \cite{Ema:2022pmo}.

We note in passing a couple of observations that reduce the amount of calculation. First, the CP-odd light-by-light operator induced by the heavy lepton loop vanishes at $ q=0 $ limit due to the Ward identity. This tells us that we need to keep $ q $ at linear order in the heavy lepton part, so $ q $ dependence on the photon propagator and light lepton part can be neglected, as long as we fix the momentum assignment of the photons. Second, not all 24 diagrams are independent. For example, the two diagrams in Fig.~\ref{equivalent diagram example} are the same (or in the general $ \text{SU(N)} $ case, differs only by a color factor). Similar relations hold for other diagrams and reduce the number of diagrams we need to calculate by a half.

Comparing with~\cite{Grozin:2008nw}, we see that their result corresponds to our $ a^{(1)} $, 
while the contribution from $ a^{(2)} $ is not included.
This means they expanded the amplitude in $ m_e/m_\tau $ but missed the expansion in $ p/m_\tau $. Our calculation shows both expansions contribute to the electron EDM at the same order. Numerically our result is $\sim 40\%$ larger than~\cite{Grozin:2008nw}. Throughout the calculation, we have assumed the ``contact" nature of the EDM source, that is no $q^2$ dependence of $d_L$ within the relevant range of momenta, $|q^2|\lesssim m_L^2$.

As a double-check, we also reevaluated the leading order contribution to electron $ g-2 $ from light-by-light scattering induced by muon loop \cite{Laporta:1992pa, Kuhn:2003pu}. The same procedure reproduces the known result correctly, and both expansions in $ m_e/m_\mu $ and in $ p/m_\mu $ need to be included to get the correct result.

With the reevaluation of heavy lepton EDM contribution to light lepton EDM, and improved experimental accuracy for $ d_e $\cite{ACME:2018yjb}, we update the constraint on tau EDM:
\begin{equation}
\label{tau EDM}
    \abs{d_\tau}<1.1\times 10^{-18}e\;\text{cm}\quad (90\% \text{C.L.}).
\end{equation}
While on the theoretical side there is only a small change from the previous work, the accuracy of paramagnetic experiments has improved by two orders of magnitude since the time \cite{Grozin:2008nw} is published, and pushed the constraint on $ d_\tau $ from $ d_e $ tighter than the one from $ e^+e^-\to\tau^+\tau^- $ by the Belle experiment.
The Belle-I\hspace{-0.1em}I experiment plans to measure $d_\tau$, again by $e^+ e^- \to \tau^+ \tau^-$, with an accuracy of $\vert\mathrm{Re}\,d_\tau\vert, \vert\mathrm{Im}\,d_\tau\vert < 10^{-18} \hyphen 10^{-19}\,e\,\mathrm{cm}$~\cite{Belle-II:2018jsg}.
Therefore the indirect constraint~\eqref{tau EDM} provides an important benchmark
for the Belle-I\hspace{-0.1em}I experiment, even though the relevant energy scale is slightly different. We note that the studies of the tau-lepton electromagnetic form factors will greatly benefit from the proposed addition of the longitudinal polarization to the electron beam \cite{Roney:2019til}, and several orders of magnitude improvements are possible \cite{Bernabeu:2007rr,Crivellin:2021spu}.

We also make an update to the indirect constraint on muon EDM derived from paramagnetic EDM experiment in~\cite{Ema:2021jds}:
\begin{equation}
\label{muon EDM}
    \abs{d_\mu}<1.7\times 10^{-20}e\;\text{cm}\quad (90\% \text{C.L.}).
\end{equation}
This is slightly better than the one provided in~\cite{Ema:2021jds}, but is not a significant change since the semi-leptonic $CP$-odd operator $ C_S $ is more important there.
In the case of tau EDM, this operator is suppressed since it scales as $C_S \propto m_L^{-3}$, and $d_e$ induced by $d_\tau$ dominates.

We can also generalize our calculation to other $CP$-odd sources. Previously we have computed \cite{Ema:2022pmo} heavy quark EDM inducing light quark EDM via the three-gluon exchange, which, apart from color factors, corresponds to diagram (a) in Fig. \ref{3-loop QED diagram}. 
Diagrams (b), (c), (d) in Figure \ref{3-loop QED diagram}, with photon propagators replaced by gluons and light/heavy leptons replaced by light/heavy quarks, also induce light quark EDMs from heavy quark CEDMs. Our result still applies here, with a slight modification to account for the color factor and the heavy quark electric charge:
\begin{equation}
\begin{aligned}
    \frac{\Delta d_q}{e}&=-\frac{Q_Q}{48}d^{abc}d^{abc}\times (a_2^{(1)}+a_2^{(2)}) \frac{m_q}{m_Q}\left(\frac{\alpha_s}{\pi}\right)^3 \tilde{d}_Q\\
    &=-\frac{5 Q_Q}{18}\times \left[\frac{11}{4}\zeta(3)-\frac{29}{24}\right] \frac{m_q}{m_Q}\left(\frac{\alpha_s}{\pi}\right)^3 \tilde{d}_Q,
\end{aligned}
\end{equation}
where $ q $ and $ Q $ stand for light and heavy quarks, respectively, $Q_Q$ is the electric charge of the heavy quark (i.e. $Q_c = 2/3$ and $Q_b = -1/3$), $d^{abc} $ is the symmetric structure constant of the $ \text{SU(3)} $ group generator, and $ \tilde{d}_Q $ is the heavy quark CEDM. For $ m_q $ and $ \alpha_s $ we use their values at the heavy quark mass scale $m_c = 1.27\,\mathrm{GeV}$ and $m_b = 4.18\,\mathrm{GeV}$, with $ m_u(m_c)=2.5\,\text{MeV} $, $ m_d(m_c)=5.4\,\text{MeV}$, $ \alpha_s(m_c)=0.38 $, $ m_u(m_b)=1.8\,\text{MeV} $, $ m_d(m_b)=4.0\,\text{MeV} $, $ \alpha_s(m_b)=0.223 $. We use $d_n=(4d_d-d_u)/3 $\cite{Pospelov:2000bw,Pospelov:2005pr,Hisano:2012sc} and $\vert d_n\vert <1.8\times 10^{-26}e\,\text{cm} $\cite{Abel:2020pzs} to put constraints on heavy quark CEDM:
\begin{equation}
\label{heavy quark CEDM}
    \begin{aligned}
        &\vert \tilde{d}_c \vert <5.2\times 10^{-21}\,\text{cm}\quad (90\% \text{C.L.}),\\
        &\vert \tilde{d}_b\vert < 2.3\times 10^{-19}\,\text{cm}\quad (90\% \text{C.L.}).
    \end{aligned}
\end{equation}
This is weaker than the constraint from the Weinberg three-gluon operator
\cite{Weinberg:1989dx,Chang:1990jv,Boyd:1990bx,Dine:1990pf,Demir:2002gg,Sala:2013osa} that will be induced by heavy quark CEDMs at one-loop order
(see {\em e.g.} \cite{Demir:2002gg,Haisch:2021hcg}).

\section{Conclusion}
In this paper we have reevaluated the three-loop heavy lepton EDM contribution 
to the electron EDM.
We have shown that expansions in electron mass and external electron momentum contribute to the tree-loop EDM diagrams at the same order in $ m_e/m_L$,
and accounting for both contributions enhances the previous result by $ 40\% $. 
With this result we have updated the indirect limits of several CP-violating observables. Eq.~(\ref{tau EDM}) shows that the constraint derived from paramagnetic EDM experiments on the tau-lepton EDM is more than an order of magnitude better than other constraints on the same form factor~\cite{Workman:2022}.  Eq.~(\ref{muon EDM}) updates the constraint on muon EDM from paramagnetic EDM experiments with a slight improvement. Eq.~(\ref{heavy quark CEDM}) provides constraints on charm and bottom quark CEDMs based on neutron EDM experiments. The constraint is weaker than the one derived from the Weinberg operator but provides a complement to the existing limit.

\paragraph{Acknowledgements}
Y.E. and M.P. are supported in part by U.S.\ Department of Energy Grant No.~de-sc0011842. Y.E. and M.P. would like to thank Perimeter Institute for the hospitality extended to them during the completion of this work.
Y.E. would like to thank Yukinari Sumino for holding a spring school on multi-loop computation at Tohoku University in 2017.
M.P. would like to thank Dr. K. Melnikov for the advice in evaluating loop contributions. 
The Feynman diagrams in this paper are generated by \texttt{TikZ-Feynman}~\cite{Ellis:2016jkw}.

\appendix

\section{Convention}
\label{app:convention}
We use the following conventions in our paper
\begin{equation}
    \begin{aligned}
        &e=|e|,\\
        &g_{\mu\nu}=\text{diag}(1,-1,-1,-1),\quad \epsilon^{0123}=+1, \quad \sigma_{\mu\nu}=\frac{i}{2}[\gamma_\mu,\gamma_\nu],\\
        &\tilde{F}^{\mu\nu}=\frac{1}{2}\epsilon^{\mu\nu\alpha\beta}F_{\alpha\beta}.
    \end{aligned}
\end{equation}
To avoid the issue of $ \gamma^5 $ in $ d $-dimension, we use the following (orthogonal but not normalized) basis for $ 4\times 4 $ gamma matrices
\begin{equation}
    \{1,\:\gamma^\mu,\:\sigma^{\mu\nu}, \:\{\sigma^{\mu\nu},\gamma^\rho\}, \:\gamma^{[\mu}\gamma^\nu\gamma^{\rho}\gamma^{\sigma]} \}.
\end{equation}
The EDM and CEDM operators are defined as
\begin{equation}
    \begin{aligned}
        &\mathcal{L}_{\text{EDM}}=\frac{d}{2}\bar{\psi}\sigma^{\mu\nu}\tilde{F}_{\mu\nu}\psi,\\
        &\mathcal{L}_{\text{CEDM}}=\frac{g_s\tilde{d}}{2}\bar{\psi}\sigma^{\mu\nu}\tilde{G}^a_{\mu\nu}T^a\psi.
    \end{aligned}
\end{equation}
The $ \text{SU(3)} $ generator $ T^a $ is normalized as
\begin{equation}
    \text{tr}_c[T^aT^b]=\frac{\delta^{ab}}{2}.
\end{equation}


\small
\bibliographystyle{utphys}
\bibliography{ref}

\end{document}